\documentclass[a4paper,12pt]{article}
\usepackage{times}
\usepackage{graphicx}
\usepackage{bm}
\newcommand{\tfrac}[2]{{\textstyle\frac{#1}{#2}}}
\newcommand{\dd}{\mbox{d}}
\newcommand{\bbox}[1]{\mbox{\boldmath $#1$}} 
\begin{document}
\title{Superconductor-Insulator Quantum Phase Transitions
\thanks{
Presented at the XL Cracow School of Theoretical Physics,
Zakopane, Poland, June 3-11, 2000.}}
\author{Adriaan M. J. Schakel \\ {\it Low Temperature Laboratory, Helsinki
University of Technology}, \\ {\it P.O. Box 2200, FIN-02015 HUT, Finland}
\\ and \\
{\it Department of Electrophysics,
National Chiao Tung University,
} \\
{\it Hsinchu, 30050, Taiwan, R.O.C.}}
\maketitle
\begin{abstract}
In these lectures, superconductivity in impure thin films close to the
absolute zero of temperature is discussed.  The behavior as function of
the applied magnetic field and the amount of impurities suggests the
presence of a superconductor-insulator transition at zero temperature.
The theory of superconductivity in the limit where all the electrons
become tightly bound in pairs is used to explain the main
characteristics of the transition.  In that limit, where the theory
becomes equivalent to a phase-only theory, electron pairs exist on
either side of the transition.  The presentation is pedagogical in
nature and includes exercises as a learning aid for those new to the
field. \\
\end{abstract}
%
%
\section{Introduction}
The topic of these lectures is superconductivity in impure thin films
close to the absolute zero of temperature.  Such superconducting films
may by increasing either the amount of impurities, or the applied
magnetic field go over to an insulating state.  The transition is
believed to signal the presence of a quantum critical point at zero
temperature \cite{MPAFisher,HPsu1}.  Quantum critical phenomena differ
from conventional critical phenomena taking place at finite temperature,
in that quantum rather than thermal fluctuations are important in the
critical region.  Since quantum phase transitions occur at zero
temperature, they cannot be induced by changing the temperature as can
thermal phase transitions, and another parameter is to be varied to
trigger the transition.  It also means that one can never tune right
through the transition as experiments are necessarily done at finite
temperature.  For low enough temperature though, the presence of a
quantum critical point can nevertheless be detected by using finite-size
scaling.

A superconducting and an insulating state seems an unlikely combination
to be present in the same system.  Whereas superconductivity needs an
{\it attractive} interaction for the pairing between electrons, the
insulating state needs a {\it repulsive} interaction.  It is a priori
not clear how these two requirements can be fulfilled in a single
system.  Adding to the bewilderment is the striking similarity in the
current-voltage characteristics in both phases close to the transition.
By interchanging the current and voltage axes in one phase, a
characteristic obtained at some value of the applied magnetic field say,
can be mapped onto a characteristic of the other phase obtained at a
different value of the field.  This means that, although the physical
properties of the superconducting and the insulating state are
completely different, there is a close connection between the conduction
mechanisms in the two phases.  A last defining aspect of the
superconductor-insulator transition is the presence of a $1/r$ Coulomb
potential, which at low charge-carrier densities becomes very strong.

In these lectures, we will argue that the main aspects of the
superconductor-insulator transition just mentioned can be accounted for
by a single theory.  Namely, the theory of superconductivity in the
limit where all the electrons become tightly bound in pairs---the
so-called {\it composite boson limit} \cite{Eagles,Leggett}.  This is
the opposite limit of the conventional weak-coupling BCS limit, where
only electrons in a thin shell around the Fermi surface become loosely
bound in Cooper pairs.  In the composite boson limit, the
superconducting state displays a quantum phase transition to an
insulating state characterized not by an unbinding of electron pairs,
but rather by a quenching of the condensate of composite bosons.  In
other words, electron pairs exist on both sides of the transition.  In
the superconducting state they are Bose-Einstein condensed, while in the
insulating state they are localized.  The attractive interaction
responsible for the binding of the electrons in pairs translates into a
repulsive interaction between the composite bosons.  The theory
describing these bosons is the Bogoliubov theory of superfluidity, which
is equivalent to a phase-only theory.  This implies that the
superconductor-insulator transition can be described by a phase-only
effective theory without ignoring any degrees of freedom
\cite{Ramakrishnan,CFGWY}.

To be fair, we should mention at this point that no consensus exists to
what extend the phase-only theory can be applied to the
superconductor-insulator transition.  Further experimental and
theoretical studies are required to settle this point.
\subsection{General Scaling}
Since quantum rather than thermal fluctuations are relevant in a quantum
phase transition, one has to work in spacetime rather than just in
space as is appropriate for thermal phase transitions in equilibrium.
As a result, in addition to a diverging correlation length $\xi$,
quantum phase transitions have also a diverging correlation time
$\xi_t$.  They indicate, respectively, the distance and time period over
which the system fluctuates coherently.  The two are related, with the
diverging correlation time scaling with the diverging correlation length
as
\begin{equation} \label{zcrit}
\xi_t \sim \xi^z,
\end{equation}
where $z$ is the so-called dynamic exponent.  It is a measure for the
asymmetry between the time and space directions close to the critical
point.  The dynamic exponent is to be added to the set of critical
exponents used to characterize a thermal phase transition.  Since such
transitions have only two independent exponents, a quantum phase
transition is specified by three independent exponents.

The traditional scaling theory of thermal, continuous phase transitions
in equilibrium, first put forward by Widom \cite{Widom}, is easily
extended to include the time dimension \cite{Ma} because the relation
(\ref{zcrit}) implies the presence of only one independent diverging
scale.

Let $\delta = K - K_{\rm c}$, with $K$ the parameter triggering the
phase transition, measure the distance from the critical value $K_{\rm
c}$.  A physical observable $O$ at the absolute zero of temperature
depends on $K$ as well as on other variable, such as an external field,
energy, or momentum.  Let us denote these other variables collectively
by $\Gamma$.  In the critical region close to the critical point, $O$
can be written as
\begin{equation} \label{scaling0}
O(\Gamma,K) = \xi^{d_O} {\cal O}(\hat{\Gamma}),
\;\;\;\;\;\;\;\; (T=0),
\end{equation}
where $d_O$ is the scaling dimension of the observable $O$, $\xi \sim
|\delta|^{- \nu}$, with $\nu$ the correlation length exponent, and
$\hat{\Gamma}$ is obtained from $\Gamma$ by rescaling it with factors of
the correlation length, so that $\hat{\Gamma}$ is independent of that
scale.  To be specific, if an external field scales as $\Gamma \sim
\xi^{d_\Gamma}$, then the rescaled field is defined as $\hat{\Gamma} =
\xi^{-d_\Gamma} \Gamma $.  The right side of Eq.\ (\ref{scaling0}) does
not depend explicitly on $K$, but only implicitly through $\xi$.  \

The data of the observable $O$ as function of an external field $\Gamma$
obtained at different values of the parameter $K$ triggering the
transition can be collapsed on a single curve when instead of
$O(\Gamma,K)$, the rescaled quantity $|\delta|^{\nu d_O} O(\Gamma,K)$ is
plotted as function not of $\Gamma$, but of $\hat{\Gamma}$.  Indeed,
because of Eq.\ (\ref{scaling0}), the combination $|\delta|^{\nu d_O}
O(\Gamma,K)$ depends only on $\hat{\Gamma}$ and is thus independent of
the distance from the critical point.  This is used to determine
critical exponents experimentally.  By rescaling the vertical and
horizontal axis of the plot with $|\delta|^x$ and $|\delta|^y$,
respectively, the best collapse obtained at some value $x_0$ and $y_0$
give the combination of critical exponents $\nu d_O = y_0$ and $\nu
d_\Gamma =x_0$.

Since a physical system is always at some finite temperature, we have to
investigate how the scaling law (\ref{scaling0}) changes when the
temperature becomes nonzero.  The easiest way to include the temperature
in a quantum theory is to go over to imaginary time $\tau = it$, with
$\tau$ restricted to the interval $0 \leq \tau \leq 1/T$.  The temporal
dimension thus becomes of finite extend.  The behavior at a finite
temperature is still controlled by the quantum critical point, provided
the correlation time satisfies $\xi_t < 1/T$.  If this condition is
fulfilled, the system will not see the finite extent of the time
dimension.  This is what makes quantum phase transitions experimentally
accessible.  Instead of the zero-temperature scaling (\ref{scaling0}),
we now have the finite-size scaling
\begin{equation} \label{scalingT}
O(\Gamma,K,T) = T^{-d_O/z} {\cal O}(\hat{\Gamma}_T, \xi_t T),
\;\;\;\;\;\;\;\; (T \neq 0),
\end{equation}
where instead of using the correlation length to convert dimensionfull
quantities in dimensionless ones, the temperature is used:
$\hat{\Gamma}_T = T^{d_\Gamma/z} \Gamma$.
\subsection*{Notation}
\label{chap:not}
We adopt Feynman's notation and denote a spacetime point by $x=x_\mu =(t,{\bf
x})$, $\mu = 0,1, \cdots,d$, with $d$ the number of space dimensions, while
the energy $k_0$ and momentum ${\bf k}$ of a particle will be denoted by
$k=k_\mu = (k_0,{\bf k})$.  The time derivative $\partial_0 =
\partial/\partial t$ and the gradient $\nabla$ are sometimes combined in a
single vector $\tilde{\partial}_\mu = (\partial_0, -\nabla)$.  The tilde on
$\partial_\mu$ is to alert the reader for the minus sign appearing in the
spatial components of this vector.  We define the scalar product $k \cdot x
= k_\mu x_\mu = k_0 t - {\bf k} \cdot {\bf x}$ and use Einstein's summation
convention.  Because of the minus sign in the definition of the vector
$\tilde{\partial}_\mu$ it follows that $\tilde{\partial}_\mu a_\mu =
\partial_0 a_0 + \nabla \cdot {\bf a}$, with $a_\mu$ an arbitrary vector.

Integrals over spacetime are denoted by
$$
\int_x = \int_{t,{\bf x}} = \int \dd t \, \dd^d x,
$$
while those over energy and momentum by
$$
\int_k = \int_{k_0,{\bf k}} = \int \frac{\dd k_0}{2 \pi}
\frac{\dd^d k}{(2 \pi)^d}.
$$
When no integration limits are indicated, the integrals are assumed to
run over all possible values of the integration variables.  Similarly,
for functional integrals we use the notation
$$
\int \mbox{D} \phi = \int_\phi.
$$ 

We will work in natural units with the speed of light, Boltzmann's
constant $k_{\rm B}$, and Planck's constant $\hbar$ set to unity.

These lectures include exercises, which are clearly marked.  Most of the
solutions can be found in Ref.\ \cite{Pompo}.
\section{Superconductivity}
In this section we study the theory of superconductivity in the limit
where all the electrons become tightly bound in pairs
\cite{Eagles,Leggett}.  The composite boson limit is to be distinguished
from the usual weak-coupling BCS limit, where only electrons (of
opposite momentum) in a thin shell around the Fermi surface become
loosely bound in Cooper pairs.
\subsection{BCS Theory}
As starting point we take the microscopic BCS model specified by the
Lagrangian \cite{BCS}
\begin{eqnarray}  \label{bcs:BCS}
     {\cal L} &=& \psi^{\ast}_{\uparrow} [i\partial_0 - \xi(-i \nabla)]
\psi_{\uparrow}  
     + \psi_{\downarrow}^{\ast} [i \partial_0 - \xi(-i
\nabla)]\psi_{\downarrow} - \lambda
\psi_{\uparrow}^{\ast}\,\psi_{\downarrow}
^{\ast}\,\psi_{\downarrow}\,\psi_{\uparrow} \nonumber \\ &:=& {\cal
L}_0 + {\cal L}_{\rm int},
\end{eqnarray} 
where ${\cal L}_0$ is the free theory, and ${\cal L}_{\rm int} = -
\lambda \psi_{\uparrow}^{\ast} \, \psi_{\downarrow}^{\ast} \,
\psi_{\downarrow}\,\psi_{\uparrow}$ is a local electron-electron
interaction term, representing the effective, phonon mediated,
attraction between electrons with coupling constant $\lambda < 0$.  The
field $\psi_{\uparrow (\downarrow )}$ is an anticommuting field
describing the electrons with mass $m$ and spin up (down), while $\xi(-i
\nabla) = \epsilon(-i \nabla) - \mu$, with $\epsilon(-i \nabla) = -
\nabla^2/2m$, is the kinetic energy operator with the chemical potential
$\mu$ characterizing the ensemble of fermions subtracted.  The theory is
invariant under global U(1) transformations under which
\begin{equation} \label{bcs:3g}
\psi_{\sigma} \rightarrow \mbox{e}^{i \alpha }
\psi_{\sigma},
\end{equation}
with $\sigma = \uparrow, \downarrow$ and $\alpha$ a constant
transformation parameter.  The superconducting state is characterized by
a spontaneous breakdown of this symmetry.

To investigate the superconducting state, we integrate out the fermionic
degrees of freedom at the expense of a newly introduced auxiliary field,
which is better equipped to describe this condensed matter state.  It is
convenient to first introduce the two-component field
\begin{equation} \label{bcs:32}
\psi = \left( \begin{array}{c} \psi_{\uparrow} \\
\psi_{\downarrow}^{\ast}
\end{array} \right) \:\:\:\:\:\: \psi^{\dagger} =
(\psi_{\uparrow}^{\ast},\psi_{\downarrow}).
\end{equation} 
In this so-called Nambu representation, ${\cal L}_0$ becomes
\begin{equation}    \label{bcs:33}
{\cal L}_0 = \psi^{\dagger}\, \left(\begin{array}{cc} i \partial_0 -
\xi(-i \nabla) & 0 \\ 0 & i \partial_0 + \xi(-i \nabla)
\end{array}\right) \, \psi,
\end{equation}
where the anticommuting character of the electron fields is used and
total derivatives are neglected.  We next replace the electron-electron
interaction with an expression involving the auxiliary field $\Delta$
\begin{equation}   \label{bcs:35} 
\lambda  \psi_{\uparrow}^{\ast} \,
\psi_{\downarrow}^{\ast} \, \psi_{\downarrow}\, \psi_{\uparrow}  \to
\Delta^* \,
\psi_{\downarrow}\,\psi_{\uparrow} + \psi_{\uparrow}^{\ast} \,
\psi_{\downarrow}^{\ast} \, \Delta - \frac{1}{\lambda} |\Delta|^2.
\end{equation} 
The original interaction is regained when the auxiliary field is
integrated out.  Physically, $\Delta$, representing a product of two
electron fields, describes electron pairs.  We shall therefore refer to
it as pair field.  With this replacement, the partition function
\begin{equation}     \label{bcs:34}
Z = \int_{\psi^{\dagger},\psi} \exp \left( i \int_x \,{\cal L} \right),
\end{equation} 
becomes
\begin{eqnarray}  \label{bcs:36}
Z = \int_{\psi^{\dagger},\psi, \Delta^*, \Delta} && \!\!\!\!
\exp\left(\frac{i}{\lambda} \int_x |\Delta|^2 \right) \\ && \!\!\!\!
\times \exp \left[ i \int_x \, \psi^{\dagger} \left(
\begin{array}{cc} i \partial_0 - \xi(-i \nabla) & -\Delta \\ -\Delta^* & i
\partial_0 + \xi(-i \nabla)
\end{array} \right)   \psi \right] \nonumber .  
\end{eqnarray}
Since the fermion fields appear quadratically now, they may be integrated
out to yield an effective action $S_{\rm eff}$ for $\Delta$ and $\Delta^*$
\begin{equation}  \label{bcs:312}
S_{\rm eff}[\Delta^*, \Delta] = -i \, {\rm Tr} \ln \left(
\begin{array}{cc} p_{0} - \xi ({\bf p}) & -\Delta \\ -\Delta^* &
p_{0} + \xi ({\bf p})
\end{array}\right),
\end{equation} 
where $p_0 = i \partial_0$ and $\xi({\bf p}) = \epsilon({\bf p}) - \mu$,
with $\epsilon({\bf p}) = {\bf p}^2/2m$ and ${\bf p} = -i \nabla$, so that
the partition function becomes
\begin{equation}   \label{bcs:37}
Z = \int _{\Delta^*, \Delta} \exp \left(i S_{\rm eff} [ \Delta^*,
\Delta] + \frac{i}{\lambda} \int_x |\Delta|^2 \right).
\end{equation}
The trace Tr appearing in Eq.\ (\ref{bcs:312}) denotes the trace over
discrete indices as well as the integration over spacetime and over
energy $k_0$ and momentum ${\bf k}$.

In the mean-field approximation, the functional integral over the pair
field in Eq.\ (\ref{bcs:37}) is approximated by the saddle point, where
only the extremal value, satisfying the equation
\begin{equation}     \label{bcs:gap}
\frac{\delta S_{\rm eff} }{\delta \Delta_0^* (x) } = - \frac{1}{\lambda}
\Delta_0(x),
\end{equation}
is included.  For a system homogeneous in spacetime, the pair field is
a constant $\bar{\Delta}$, and Eq.\ (\ref{bcs:gap}) reduces, after passing
to the Fourier representation, to the BCS gap equation \cite{BCS}:
\begin{eqnarray}   \label{bcs:gape} 
\frac{1}{\lambda} &=& - i  \int_k \frac{1}{k_0^{2} - E^{2}(k) + i \eta}
\nonumber \\ &=& - \frac{1}{2} \int_{\bf k} \frac{1}{E({\bf k})}.
\end{eqnarray} 
Here, $\eta$ is an infinitesimal positive constant which is to be set to
zero at the end of the calculation, and 
\begin{equation}  \label{bcs:spec}
E({\bf k}) = \sqrt{\xi^2({\bf k}) + |\bar{\Delta}_0|^2}
\end{equation}  
is the spectrum\footnote{To avoid confusion, let us repeat that the bar
in $\bar{\Delta}_0$ indicates that the pair field is a constant,
while the subscript $0$ indicates that it satisfies the extremal
condition (\ref{bcs:gap}).} of the fermionic excitations.  A nontrivial
solution to the gap equation signals the spontaneous symmetry breakdown
of the global U(1) symmetry (\ref{bcs:3g}).
\subsection{Composite Boson Limit}
For studying the composite boson limit, it proves prudent to swap the
coupling constant $\lambda$ in the gap equation (\ref{bcs:gape}) for a
more convenient parameter, namely the binding energy $\epsilon_a$ of an
electron pair in vacuum \cite{RDS}.  Both parameters characterize the
strength of the electron-electron interaction.  To establish the
connection between the two, let us consider the Schr\"odinger equation
for the problem at hand.

In reduced coordinates, it reads
\begin{equation} 
\left[- \frac{\nabla^2}{m} + \lambda \, \delta({\bf x}) \right] \psi({\bf
x}) = - \epsilon_a,
\end{equation} 
where the reduced mass is $m/2$ and the delta-function potential, with
$\lambda < 0$, represents the attractive local electron-electron
interaction ${\cal L}_{\rm int}$ in Eq.\ (\ref{bcs:BCS}).  We stress
that this is a two-particle problem in vacuum and not the famous Cooper
problem of two interacting fermions on top of a filled Fermi sea.  The
equation is most easily solved in the Fourier representation, yielding
\begin{equation} 
\psi({\bf k}) = - \frac{\lambda}{{\bf k}^2/m + \epsilon_a} \psi(0),
\end{equation} 
or
\begin{equation} \label{bound}
- \frac{1}{\lambda} = \int_{\bf k} \frac{1}{{\bf k}^2/m + \epsilon_a} .
\end{equation} 
This bound-state equation allows us to replace the coupling constant
$\lambda$ with the binding energy $\epsilon_a$.  When substituted in
the gap equation (\ref{bcs:gape}), the latter becomes
\begin{equation} \label{bcs:reggap}
\int_{\bf k} \frac{1}{{\bf k}^2/m + \epsilon_a} = \frac{1}{2}
\int_{\bf k} \frac{1}{E({\bf k})}.
\end{equation} 
By inspection, it follows that this equation has a solution given by
\cite{Eagles,Leggett}
\begin{equation} \label{comp:self}
\bar{\Delta}_0 \rightarrow 0, \;\;\;\;\; \mu \rightarrow - \tfrac{1}{2}
\epsilon_a,
\end{equation}   
where it should be noted that, in contrast to the weak-coupling limit,
the chemical potential characterizing the ensemble of fermions is {\it
negative} here.  This is the composite boson limit.

To appreciate the physical significance of the specific value found for
the chemical potential in this limit and also its name, observe that the
spectrum $E_{\rm B}({\bf q})$ of the two-fermion bound state measured
relative to the pair chemical potential $2\mu$ reads
\begin{equation} 
E_{\rm B}({\bf q}) = \frac{{\bf q}^2}{4m} - \mu_{\rm B},
\end{equation} 
where $\mu_{\rm B}$ is defined as $\mu_{\rm B} = \epsilon_a + 2 \mu$
and may be understood as the chemical potential characterizing the
ensemble of composite bosons.  The negative value for $\mu$ found in
Eq.\ (\ref{comp:self}) is precisely the condition for a Bose-Einstein
condensation of an {\it ideal} gas of composite bosons in the ${\bf q} =
0$ state.  

Including quadratic terms in $\bar{\Delta}_0$, we obtain as solution to
Eq.\ (\ref{bcs:reggap})
\begin{equation} 
\mu = - \frac{1}{2} \epsilon_a + (1-d/4)
\frac{|\bar{\Delta}_0|^2}{\epsilon_a}.
\end{equation} 
This leads to the chemical potential 
\begin{equation} \label{chem}
\mu_{\rm B} = (2-d/2) \frac{|\bar{\Delta}_0|^2}{\epsilon_a},
\end{equation} 
characterizing the now interacting Bose gas.
\subsection{Renormalization}
\label{sec:ren}
For a system homogeneous in spacetime, so that the field $\Delta_0(x)$
is constant, the effective action (\ref{bcs:312}) is readily evaluated.
Disassembling the argument of the logarithm as
\begin{equation} 
\left(
\begin{array}{cc} p_{0} - \xi ({\bf p}) & -\bar{\Delta}_0 \\
-\bar{\Delta}_0^* & p_{0} + \xi ({\bf p}) \end{array}\right) = \left(
\begin{array}{cc} p_{0} - \xi ({\bf p}) & 0 \\ 0 &
p_{0} + \xi ({\bf p}) \end{array}\right) 
  - \left(
\begin{array}{cc} 0 & \bar{\Delta}_0 \\ \bar{\Delta}_0^* & 0
\end{array}\right), 
\end{equation} 
and expanding the second logarithm in a Taylor series, we recognize the
form
\begin{eqnarray}  \label{resum}
S_{\rm eff}[\bar{\Delta}_0^*, \bar{\Delta}_0] = &&   -i \,
{\rm Tr} \ln \left(
\begin{array}{cc} p_{0} - \xi ({\bf p}) & 0 \\ 0 &
p_{0} + \xi ({\bf p}) \end{array}\right) \nonumber \\ && 
 - i \, {\rm Tr}
\ln \left(1 - \frac{|\bar{\Delta}_0|^2}{p_0^2 - \xi^2({\bf p})} \right),
\end{eqnarray}  
apart from an irrelevant constant.  Again passing to the Fourier
representation, and carrying out the integral over the loop energy
$k_0$, we obtain the effective Lagrangian
\begin{equation} \label{Exi}
{\cal L}_{\rm eff} = \int_{\bf k} \left[ E({\bf k}) - \xi({\bf k})
\right].
\end{equation} 

\hrule  \vskip1pt \hrule \vskip4pt
{\footnotesize
\noindent {\bf Exercise:} Derive this result from Eq.\ (\ref{resum}),
using contour integration.  Rather than integrating the logarithms in
that equation, one better first differentiate it with respect to the
chemical potential as the integral over the loop energy $k_0$ becomes
easier that way.  In the end one integrates the resulting expression
again over the chemical potential to obtain the desired result
(\ref{Exi}).} 
\vskip2pt \hrule  \vskip1pt \hrule \vskip4pt

To the one-loop result (\ref{Exi}), the tree term
$|\bar{\Delta}_0|^2/\lambda$ is to be added.  Expanding $E({\bf k})$ in
Eq.\ (\ref{Exi}) in a Taylor series, we see that the effective
Lagrangian also contains a term quadratic in $\bar{\Delta}_0$.  This
term amounts to a renormalization of the coupling constant.
Specifically, the renormalized coupling constant $\lambda_{\rm r}$ is to
this order given by
\begin{eqnarray}  \label{bcs:reng}
\frac{1}{\lambda_{\rm r}} &=& \frac{1}{\lambda} + \frac{1}{2} \int_{\bf
k} \frac{1}{\xi({\bf k})} \nonumber \\ &=& \frac{1}{\lambda} +
\frac{\Gamma(1-d/2)}{(4 \pi)^{d/2}} \frac{m^{d/2}}{\epsilon_a^{1-d/2}},
\end{eqnarray} 
with, as is appropriate in the composite boson limit,
\begin{equation} \label{approx} 
\xi({\bf k}) =  \epsilon({\bf k}) + \tfrac{1}{2} \epsilon_a, 
\end{equation} 
and dimensional regularization is used in evaluating the momentum
integral.  Because of Eq.\ (\ref{bound}), which may be viewed as the
defining equation of the parameter $\epsilon_a$, the right side of Eq.\
(\ref{bcs:reng}) is zero.  This implies that in the composite boson
limit, $\lambda_{\rm r} \to - \infty$ so that we indeed have tightly
bound pairs here.
\section{Bogoliubov theory}
In this section we show that in the composite boson limit, the effective
theory obtained after integrating out the fermionic degrees of freedom,
is the Bogoliubov theory of superfluidity in an interacting Bose gas.
The gas consists of composite bosons with a mass twice the electron
mass.  The system is known to undergo a quantum phase transition from
the superfluid to a (Mott) insulating state \cite{FWGF}.  When
translated back to the fermionic theory, this transition corresponds to
one where the condensate is drained of composite bosons, without
breaking them up.  We include impurities in the Bogoliubov theory to
show that it leads to localization without destroying the superfluid
state completely.  The insulating state, which is now a result not only
of repulsive interactions, as in a Mott insulator, but also of
(Anderson) localization, is called an Anderson-Mott insulator.
\subsection{Derivative Expansion}
We next wish to relax the assumption of homogeneity in spacetime and
consider a spacetime-dependent pair field.  To this end, we study
the effective action (\ref{bcs:312}) and expand $\Delta(x)$ around the
constant value $\bar{\Delta}_0$ satisfying the gap equation
(\ref{bcs:gape}),
\begin{equation} 
\Delta(x) = \bar{\Delta}_0 + \tilde{\Delta}(x).
\end{equation} 
We obtain in this way,
\begin{equation} \label{effact}
S_{\rm eff} = i \, {\rm Tr} \sum_{\ell =1}^\infty \frac{1}{\ell} \left[
G_0(p) \left( \begin{array}{cc} 0 & \tilde{\Delta} \\ \tilde{\Delta}^* &
0 \end{array} \right) \right]^\ell,
\end{equation} 
where $G_0$ is the correlation function,
\begin{eqnarray}    \label{bcs:prop}
G_0(k) &=&
\left( \begin{array}{cc} k_0 - \xi  ({\bf k}) 
& -\bar{\Delta}_0 \\ -\bar{\Delta}_0^*  & k_0 + \xi ({\bf k}) 
\end{array} \right)^{-1}  \\ &=& 
\frac{1}{k_0^2 - E^2({\bf k}) + i  \eta } 
\left( \begin{array}{cc} k_0 \, {\rm e}^{i k_0 \eta } + \xi
({\bf k})  & 
\bar{\Delta}_0 \\ \bar{\Delta}_0^* & k_{0} \, {\rm e}^{-i k_0 \eta}- \xi
({\bf k}) \end{array} \right). \nonumber 
\end{eqnarray}
The exponential functions in the diagonal elements of the correlation
function are  additional convergence factors needed in nonrelativistic
theories \cite{Mattuck}.

When evaluating the effective action (\ref{effact}), the precise meaning
of the trace Tr appearing there should be kept in mind.  Explicitly, it
is defined as
\begin{equation}   \label{bcs:explicit}
S_{\rm eff} = -i {\rm Tr} \, \ln \left[K(p,x) \right] = -i {\rm tr}
\ln\left[ K(p,x) \delta (x - y)\bigr|_{y = x} \right],
\end{equation}
where the trace tr is the usual one over discrete indices.  We
abbreviated the matrix appearing in Eq.\ (\ref{bcs:312}) by $K(p,x)$ so
as to cover the entire class of actions of the form
\begin{equation} 
S = \int_x \psi^\dagger(x) K(p,x) \psi(x).
\end{equation} 
The delta function in Eq.\ (\ref{bcs:explicit}) arises because $K(p,x)$
is obtained as a second functional derivative of the action
\begin{equation}  
\frac{\delta^{2} S}{\delta \psi^\dagger(x) \, \delta \psi(x)} =
K(p,x)  \,  \delta (x - y) \bigr|_{y =  x},
\end{equation} 
each of which gives a delta function.  Since the action contains only
one integral over spacetime, one delta function remains.  Because it is
diagonal, the delta function may be taken out of the logarithm and
Eq.\ (\ref{bcs:explicit}) can be written as
\begin{eqnarray}  \label{bcs:Trexplicit}
S_{\rm eff} &=& -i {\rm tr} \, \int_x 
\ln \left[ K(p,x) \right] 
\delta (x - y) \bigr|_{y = x}  \nonumber  \\ &=&
-i {\rm tr} \, \int_x \int_k {\rm e}^{i k \cdot x} \, \ln \left[ K(p,x)
\right]  {\rm e}^{-i k \cdot x}.
\end{eqnarray}
In the last step, we used the integral representation of the
delta function:
\begin{equation}
\delta (x) = \int_k {\rm e}^{-i k \cdot x},
\end{equation}
shifted the exponential function $\exp (i k \cdot y)$ to the left, which
is justified because the derivative $p_\mu$ does not operate on it, and,
finally, set $y_\mu$ equal to $x_\mu$.  We thus see that the trace Tr in
Eq.\ (\ref{bcs:explicit}) stands for the trace over discrete indices as
well as the integration over spacetime and over energy and momentum.
The integral $\int_k$ arises because the effective action calculated
here is a one-loop result with $k_\mu$ the loop energy and momentum.

The integrals in Eq.\ (\ref{bcs:Trexplicit}) cannot in general be
evaluated in closed form because the logarithm contains energy-momentum
operators and spacetime-dependent functions in a mixed order.  To
disentangle the integrals resort has to be taken to a derivative
expansion \cite{FAF} in which the logarithm is expanded in a Taylor
series.  Each term contains powers of the energy-momentum operator
$p_\mu$ which acts on every spacetime-dependent function to its right.
All these operators are shifted to the left by repeatedly applying the
identity
\begin{equation} 
f(x) p_\mu g(x) = (p_\mu - i \tilde{\partial}_\mu) f(x) g(x),
\end{equation} 
where $f(x)$ and $g(x)$ are arbitrary functions of spacetime and the
derivative $\tilde{\partial}_\mu = (\partial_0,-\nabla)$ acts {\it only} on
the next object to the right.  One then integrates by parts, so that all the
$p_\mu$'s act to the left where only a factor $\exp(i k \cdot x)$ stands.
Ignoring total derivatives and taking into account the minus signs that
arise when integrating by parts, one sees that all occurrences of $p_\mu$
(an operator) are replaced with $k_\mu$ (an integration variable).  The
exponential function $\exp(-i k \cdot x)$ can at this stage be moved to the
left where it is annihilated by the function $\exp(i k \cdot x)$.  The
energy-momentum integration can now in principle be carried out and the
effective action be cast in the form of an integral over a local density
${\cal L}_{\rm eff}$:
\begin{equation}   
S_{\rm eff} = \int_x {\cal L}_{\rm eff}.
\end{equation} 
This is in a nutshell how the derivative expansion works. 

\vskip4pt \hrule \vskip1pt \hrule \vskip4pt 
{\footnotesize 
\noindent {\bf Exercise:} Apply the derivative expansion to the
Lagrangian
\begin{equation} 
{\cal L} = \tfrac{1}{2} (\tilde{\partial}_\mu \phi)^2 - \tfrac{1}{2} m^2
\phi^2 - \tfrac{1}{4} \lambda \phi^4,
\end{equation}
with $\phi$ a real scalar field.  The theory has a Z$_2$ symmetry under
which the scalar field flips sign: $\phi \to \phi' = - \phi$.  Show that
the effective theory in two space dimensions is given by
\begin{equation} 
{\cal L}_{\rm eff} = \tfrac{1}{2} Z(\phi) (\tilde{\partial}_\mu \phi)^2
- {\cal V}_{\rm eff}(\phi),
\end{equation} 
with 
\begin{eqnarray} 
{\cal V}_{\rm eff}(\phi) &=& - \frac{1}{12 \pi} (m^2 + \tfrac{1}{2}
\lambda \phi^2)^{3/2}, \nonumber \\ {\cal Z}(\phi) &=& \frac{1}{192 \pi}
\frac{\lambda^2 \phi^2}{(m^2 + \tfrac{1}{2} \lambda \phi^2)^{3/2}}.
\end{eqnarray} 
}
\vskip2pt \hrule  \vskip1pt \hrule \vskip4pt
\subsection{Map onto Bogoliubov Theory}
We are interested in terms in the effective action (\ref{effact}) quadratic
in the field $\tilde{\Delta}$.  Using the derivative expansion, we find
\begin{eqnarray}  \label{comp:Seff}
S_{\rm eff}^{(2)}(q)  &=&  \frac{i}{2} \, {\rm Tr} \,
\frac{1}{p_0^2 - E^2({\bf p})} 
\frac{1}{(p_0 + q_0)^2 - E^2({\bf p} - {\bf q})}  \\ &&
\;\;\;\; \times 
\Bigr\{ \bar{\Delta}_0^2 \, \tilde{\Delta}^* \tilde{\Delta}^*  
+ [p_0 + \xi({\bf p})] [p_0 + q_0 - \xi({\bf p} - {\bf q})] \tilde{\Delta}
\tilde{\Delta}^* \nonumber \\ && \;\;\;\;\;\;\;\;\; +
\bar{\Delta}_0^{*^{\scriptstyle{2}}} 
\tilde{\Delta} \tilde{\Delta}  
+ [p_0 - \xi({\bf p})] [p_0 + q_0 + \xi({\bf p} - {\bf q})] \tilde{\Delta}^*
\tilde{\Delta} \Bigl\}, \nonumber 
\end{eqnarray} 
where $q_\mu = i\tilde{\partial}_\mu$.  Let us first ignore the
derivatives in this expression.  After carrying out the integral over
the loop energy $k_0$ and using the gap equation (\ref{bcs:gape}), we
obtain
\begin{equation} \label{comp:Lag1}
{\cal L}^{(2)}(0) = -\frac{1}{8} \int_{\bf k} \frac{1}{E^3({\bf k})}
\left(\bar{\Delta}_0^2 \, \tilde{\Delta}^*{}^2 +
\bar{\Delta}_0^{*^{\scriptstyle{2}}}  \tilde{\Delta}^2 + 2 
|\bar{\Delta}_0|^2 |\tilde{\Delta}|^2 \right).
\end{equation} 
In the composite boson limit, where the spectrum of the fermionic
excitations is given by Eq.\ (\ref{approx}), the integral over the loop
momentum becomes elementary, yielding
\begin{equation}
\int_{\bf k} \frac{1}{E^3({\bf k})} = \frac{4 \Gamma(3-d/2)}{(4 \pi)^{d/2}}
m^{d/2} \epsilon_a^{d/2-3}.
\end{equation} 

We next consider the terms in Eq.\ (\ref{comp:Seff}) involving
derivatives.  Following Ref.\ \cite{Haussmann} we set $\bar{\Delta}_0$
to zero here.  The integral over the loop energy is easily carried out,
with the result
\begin{eqnarray} 
{\cal L}^{(2)}(q) &=& - \frac{1}{2} \int_{\bf k}
\frac{1}{q_0 - {\bf k}^2/m + 
2 \mu - {\bf q}^2/4m} \tilde{\Delta} \tilde{\Delta}^* \nonumber \\ && 
 - \frac{1}{2} \int_{\bf k} \frac{1}{-q_0 - {\bf k}^2/m +
2 \mu - {\bf q}^2/4m} \tilde{\Delta}^* \tilde{\Delta}.
\end{eqnarray} 
In the composite boson limit, the remaining momentum integrals become
elementary again and after expanding in derivatives we find
\begin{eqnarray} \label{derivatives}
\lefteqn{\int_{\bf k} \frac{1}{q_0 - {\bf k}^2/m - \epsilon_a - {\bf
q}^2/4m} =} \\ && 
- \frac{ \Gamma(1-d/2)}{(4 \pi)^{d/2}} m^{d/2} \epsilon_a^{d/2-1} -
\frac{ \Gamma(2-d/2)}{(4 \pi)^{d/2}} m^{d/2}
\epsilon_a^{d/2-2} \left(q_0 - \frac{{\bf q}^2}{4m} \right).
\nonumber 
\end{eqnarray}  
When combined with the tree term $|\tilde{\Delta}|^2/\lambda$, the first
term at the right side of this equation yields the renormalization
(\ref{bcs:reng}) of the coupling constant.  The second term at the right
side of Eq.\ (\ref{derivatives}) gives, when combined with the
contribution (\ref{comp:Lag1}), the result \cite{Haussmann},
\begin{equation} \label{fin}
{\cal L}^{(2)} = \frac{1}{2} \frac{\Gamma(2-d/2)}{(4 \pi)^{d/2}} m^{d/2}
\epsilon_a^{d/2-2}\, \tilde{\Psi}^\dagger \, 
M_0(q) \, \tilde{\Psi}, \;\;\;\;\;  \tilde{\Psi} = \left(\begin{array}{l}
\tilde{\Delta} \\ \tilde{\Delta}^* \end{array} \right),
\end{equation} 
where $M_0(q)$ stands for the $2 \times 2$ matrix,
\begin{eqnarray}    \label{comp:M} 
\lefteqn{M_0(q) =} \nonumber \\ && 
\left( \begin{array}{cc}
q_0 - {\bf q}^2/4m - (2-d/2) |\bar{\Delta}_0|^2/ \epsilon_a & 
 - (2-d/2) \bar{\Delta}_0^2/ \epsilon_a   \\
- (2-d/2) \bar{\Delta}_0^{*^{\scriptstyle{2}}} / \epsilon_a
&  -q_0 - {\bf q}^2/4m - (2-d/2) |\bar{\Delta}_0|^2/ \epsilon_a
\end{array} \right). \nonumber \\ && 
\end{eqnarray} 
As we shall show now, this is the Bogoliubov theory of superfluidity in
an interacting Bose gas.  That is to say, after integrating out the
fermionic degrees of freedom from the theory of superconductivity, we
obtain in the composite boson limit a theory describing a gas of
repulsively interacting (composite) bosons.
\subsection{Quantum Phase Transition}
The Bogoliubov theory is specified by the Lagrangian \cite{Bogoliubov}
\begin{equation} \label{bec:Lagr}
{\cal L} = \phi^* \bigl[i \partial_0 - \epsilon(-i \nabla) + \mu_{\rm B}
\bigr] \phi - \lambda_{\rm B} |\phi|^4,
\end{equation} 
where $\mu_{\rm B}$ is the chemical potential characterizing the Bose
gas.  The self-coupling is taken to be positive, $\lambda_{\rm B} > 0$,
so that the local interaction is repulsive.

At the mean-field, or classical level, where quantum fluctuations are
ignored, the theory (\ref{bec:Lagr}) undergoes a phase transition when
the chemical potential changes sign.  For $\mu_{\rm B}>0$, the global
U(1) symmetry is spontaneously broken by a nontrivial ground state,
while for $\mu_{\rm B}<0$ the symmetry is unbroken.  The change in
$\mu_{\rm B}$ can be induced by varying the temperature, as in a thermal
phase transition, but it can also be induced at zero temperature by
varying, for example, the number of charge carriers, or the amount of
impurities.  The zero-temperature quantum phase transition describes a
transition between a superfluid and an insulating state \cite{FWGF}.

The ground state of a system homogeneous in spacetime is obtained by
considering the shape of the potential energy
\begin{equation} \label{bec:V}
{\cal V} = - \mu_{\rm B} |\bar{\phi}|^2 + \lambda_{\rm B} |{\bar
\phi}|^4.
\end{equation} 
For $\mu_{\rm B} > 0$ it indeed has a minimum away from the origin $\phi
= 0$ given by
\begin{equation}  \label{bec:min}
|\bar{\phi}_0|^2 = \frac{1}{2} \frac{\mu_{\rm B}}{\lambda_{\rm B} },
\end{equation}  
and the potential becomes
\begin{equation}  \label{V0}
{\cal V}_0 = - \frac{\mu_{\rm B}^2}{4 \lambda_{\rm B}}.
\end{equation} 
Since the total particle number density $n_{\rm B}$ is represented by
\begin{equation} 
n_{\rm B}(x) = |\phi(x)|^2,
\end{equation} 
the quantity $n_0 := |\phi_0|^2$ physically denotes the number density
of particles residing in the ground state.  A nonzero value for $n_0$
thus signals Bose-Einstein condensation.  For a homogeneous system in
its ground state, we see that at the mean-field level $\bar{n}_0 =
\bar{n}$ so that all the particles reside in the condensate.  This will
change when quantum fluctuations are included as a result of which
particles are knocked out of the condensate (see below).
	
To account for the nontrivial ground state, we introduce the shifted
field\footnote{Similar as before, the bar in $\bar{\phi}_0$ denotes a
constant value of the field, while the subscript $0$ indicates that it
satisfies the mean-field equation (\ref{bec:min}).} $\tilde{\phi}(x)$:
\begin{equation}  \label{bec:newfields}
\phi(x) = \bar{\phi}_0 + \tilde{\phi}(x).
\end{equation}
The terms in the Lagrangian (\ref{bec:Lagr}) quadratic in this shifted
field may be cast in the matrix form
\begin{equation}  \label{bec:L0}
{\cal L}_0 = \tfrac{1}{2} \tilde{\Phi}^{\dagger} M_0(p) \tilde{\Phi},
\;\;\;\;\;\; \tilde{\Phi} = \left(\begin{array}{l} \tilde{\phi} \\
\tilde{\phi}^* \end{array} \right),
\end{equation}
with
\begin{equation}   \label{bec:M} 
M_0(p) = \left( \begin{array}{cc} p_0 - \epsilon({\bf p}) + \mu_{\rm B}
- 4 \lambda_{\rm B} |\bar{\phi}_0|^2 & - 2 \lambda_{\rm B}
\bar{\phi}_0^2 \\ - 2 \lambda_{\rm B} \bar{\phi}_0^{*^{\scriptstyle{2}}} &
-p_0 - \epsilon ({\bf p}) + \mu_{\rm B} - 4 \lambda_{\rm B}
|\bar{\phi}_0|^2
\end{array} \right).
\end{equation} 
Taking into account only the quadratic terms in the field and neglecting
higher-order terms, as we just did, is known as the Bogoliubov approximation.

Comparing this expression with Eq.\ (\ref{comp:M}) obtained in the
composite boson limit after integrating out the fermionic degrees of
freedom from the theory of superconductivity, we conclude that the
composite bosons have---as expected---a mass $m_{\rm B}=2m$ twice the
fermion mass $m$, and a small chemical potential given by Eq.\
(\ref{chem}), which we there derived from the gap equation.  It also
follows that the number density of composite bosons condensed in the ground
state reads
\begin{equation} 
\bar{n}_0 = \frac{\Gamma(2-d/2)}{(4 \pi)^{d/2}} m^{d/2}
\epsilon_a^{d/2-2} |\bar{\Delta}_0|^2,
\end{equation}
and that the interaction $\lambda_{\rm B}$ between the composite bosons
is
\begin{equation} \label{comp:lambda}
\lambda_{\rm B} = (4 \pi)^{d/2} \frac{1-d/4}{\Gamma(2-d/2)}
\frac{\epsilon_a^{1-d/2}}{m^{d/2}},
\end{equation}
or, using Eq.\ (\ref{bound}),
\begin{equation} 
\lambda_{\rm B} = - \frac{1-d/4}{1-d/2} \lambda .
\end{equation} 
Note that the parameter $\lambda (<0)$ characterizing the attractive
electron-electron interaction appears below $d=2$ with a minus sign
here, leading to a repulsive interaction between the composite bosons.
(In the next subsection, we will see that $d=2$ is the upper critical
dimension of the $T=0$ Bogoliubov theory.)  This brings us to the
important conclusion that for $d<2$ the same interaction responsible for
the formation of electron pairs, is also responsible for the stability
of the superfluid state, and when this state ceases to exist, for that
of the insulating state, which both need a repulsive interaction.

The quantum phase transition encoded in the Bogoliubov theory
corresponds, when translated back to the fermionic theory, to one where
the condensate is drained of composite bosons, without breaking them up.
In other words, composite bosons exist on both sides of the transition,
either condensed (superfluid state) or localized (insulating state)
\cite{Ramakrishnan,CFGWY}.

\subsection{Beyond mean-field theory}
\label{sec:beyond}
We can continue now and improve on the usual mean-field approximation of
the theory of superconductivity, where the functional integral over the
pair field in the partition function (\ref{bcs:37}) is approximated by
the saddle point, by integrating out the field $\tilde{\Psi}$ in Eq.\
(\ref{fin}), or to simplify notation, the field $\tilde{\Phi}$ in Eq.\
(\ref{bec:L0}).  This leads to the effective potential
\begin{equation} \label{eff:Veff}
{\cal V}_{\rm eff} = -\frac{i}{2} {\rm tr} \int_k \ln[M_0(k)] =
\frac{1}{2} \int_{\bf k} E({\bf k}).
\end{equation}
Here, $E({\bf k})$ is the famous single-particle Bogoliubov spectrum
\cite{Bogoliubov},
\begin{eqnarray}  \label{eff:bogo}
E({\bf k}) &=& \sqrt{ \epsilon ^2({\bf k}) + 2 \mu_{\rm B} \epsilon({\bf
k}) } \nonumber \\ &=& \sqrt{ \epsilon ^2({\bf k}) + 4 \lambda_{\rm B}
|\bar{\phi}_0|^2 \epsilon({\bf k}) }.
\end{eqnarray}
In the limit of large momentum, the spectrum behaves in a way
\begin{equation} \label{eff:med}
E({\bf k}) \sim \epsilon({\bf k}) + 2 \lambda_{\rm B} |\bar{\phi}_0|^2
\end{equation}
typical for a nonrelativistic particle of mass $m$ moving in a
background medium, provided by the condensate in this case.  The most
notable feature of the Bogoliubov spectrum is that it is gapless,
behaving for small momentum as
\begin{equation} \label{eff:micror}
E({\bf k}) \sim c \, |{\bf k}|,
\end{equation}
with $c = \sqrt{\mu_{\rm B}/m}$.

\vskip4pt \hrule  \vskip1pt \hrule \vskip4pt
{\footnotesize 
\noindent {\bf Exercise:} Carry out the integral over the loop energy
$k_0$ in Eq.\ (\ref{eff:Veff}) using contour integration and show that
this leads to the right side of that equation.  This is best done by
first differentiating the expression with respect to the chemical
potential $\mu_{\rm B}$, and in the end integrating the result again
with respect to $\mu_{\rm B}$.}  
\vskip2pt \hrule  \vskip1pt \hrule \vskip4pt

The integral over the loop momentum in Eq.\ (\ref{eff:Veff}) can be
carried out using the integral representation of the Gamma function
\begin{equation}  \label{gamma}
\frac{1}{a^z} = \frac{1}{\Gamma(z)} \int_0^\infty \frac{\dd
\tau}{\tau} \tau^z {\rm e}^{-a \tau}.
\end{equation}
In arbitrary space dimension $d$ this yields, using dimensional
regularization:
\begin{equation} \label{regularized}
{\cal V}_{\rm eff} = - L_d m^{d/2} \mu_{\rm B}^{d/2 + 1}, \;\;\; L_d =
\frac{\Gamma(1-d/2) \Gamma(d/2 + 1/2)}{2 \pi^{d/2 + 1/2} \Gamma(d/2+2)}.
\end{equation}
For $d=2$, the effective potential diverges.  To investigate this, we
set $d=2- \varepsilon$, with $\epsilon$ small and positive, and expand
${\cal V}_{\rm eff}$ around $d=2$, giving
\begin{equation}  \label{Vep}
{\cal V}_{\rm eff} = - \frac{m}{4 \pi \varepsilon} \frac{\mu_{\rm
B}^2}{\kappa^{\varepsilon/2}} ,
\end{equation} 
with $\kappa$ an arbitrary renormalization group scale parameter which
enters for dimensional reasons.  If the Bogoliubov spectrum had not been
gapless, but had an energy gap instead, this parameter would have
appeared in Eq.\ (\ref{Vep}) in the place of $\kappa$.  As always in
dimensional regularization, the divergence shows up as a pole in
$\epsilon$.  Comparing the one-loop contribution with the classical
contribution (\ref{V0}), we conclude that Eq.\ (\ref{Vep}) leads to a
renormalization of the coupling constant $\lambda$, yielding the
renormalized coupling $\lambda_{\rm r}$ \cite{Uzunov}
\begin{equation} 
\frac{1}{\hat{\lambda}_{\rm r}} = \frac{1}{\hat{\lambda}} +
\frac{m}{\pi} \frac{1}{\varepsilon},
\end{equation} 
where $\hat{\lambda} = \lambda /\kappa^{\varepsilon/2}$ and a similar
definition for $\hat{\lambda}_{\rm r}$.  The quantum critical point is
approached by letting the renormalized group scale parameter $\kappa \to
0$.  For fixed coupling $\lambda$, it then follows that upon approaching
the critical point, the renormalized coupling tends to
$\hat{\lambda}_{\rm r} \to \pi \varepsilon /m$.  For $d<2$, or
equivalently $\epsilon >0$, the fixed point is nontrivial.  In the limit
$d\to 2$, $\hat{\lambda}_{\rm r}\to 0$ and the theory becomes Gaussian,
identifying $d=2$ as the upper critical dimension.

Due to quantum fluctuations not all the particles are known to reside in
the condensate \cite{FW}.  Specifically, in $d$ space dimensions, the
(constant) particle number density $\bar{n}$ at the one-loop level is
given by \cite{TN}
\begin{equation} \label{depletion}
\bar{n} = |\bar{\phi}_0|^2 - 2^{d/2-2} \frac{d^2-4}{d-1} L_d m^{d/2}
\lambda_{\rm B}^{d/2} |\bar{\phi}_0|^d. 
\end{equation} 
Since the quantum-induced term is positive for $1<d<4$, the number of
particles residing in the condensate given $\bar{n}$ is reduced compared
to the classical result $\bar{n} = |\bar{\phi}_0|^2$. This shows that
due to quantum fluctuations, particles are knocked out of the
condensate.

\vskip4pt \hrule  \vskip1pt \hrule \vskip4pt
{\footnotesize 
\noindent {\bf Exercise:} Derive Eq.\ (\ref{depletion}).  In doing so,
one should not use the mean-field equation (\ref{bec:min}) too early,
and instead work with the more general single-particle spectrum
\begin{equation} \label{bec:bogog}
E({\bf k}) = \sqrt{\bigl[ \epsilon({\bf k})  - \mu_0 + 4 \lambda_0
|\bar{\phi}|^2 \bigr]^2 - 4 \lambda_0^2 |\bar{\phi}|^4 } \, .
\end{equation}
It reduces to the Bogoliubov spectrum when the mean-field equation is
used. } 
\vskip2pt \hrule  \vskip1pt \hrule \vskip4pt

Despite that due to quantum fluctuations not all the particles reside in
the condensate, all the particles do in the absence of impurities and at
zero temperature participate in the superflow, and move on the average
with the superfluid velocity.  Put differently, the superfluid mass
density $\rho_{\rm s}$ is given by the total particle number density
$n$: $\rho_{\rm s} = m n$.

\vskip4pt \hrule  \vskip1pt \hrule \vskip4pt
{\footnotesize 
\noindent {\bf Exercise:} Prove this.  To this end, assume that the
entire system moves with a velocity ${\bf u}$ relative to the laboratory
system.  As in standard hydrodynamics, the time derivative in the frame
following the motion of the fluid is $\partial_0 + {\bf u} \cdot
\nabla$.  Also assume that the condensate moves with the superfluid
velocity ${\bf v}_{\rm s}$ and boost the field:
\begin{equation} 
\phi (x) \rightarrow \phi'(x) = {\rm e}^{im {\bf v}_{\rm s} \cdot {\bf x}}
\phi (x) .
\end{equation} 
Show that when incorporated in the Lagrangian (\ref{bec:Lagr}) of the
interacting Bose gas, these two changes result in a change of the
chemical potential
\begin{equation} \label{bec:mureplacement}
\mu_{\rm B} \rightarrow \mu_{\rm eff} := \mu_{\rm B} - \tfrac{1}{2} m {\bf v}_{\rm s}
\cdot ({\bf v}_{\rm s} - 2 {\bf u}) .
\end{equation} 
Show that the resulting Bogoliubov spectrum and thermodynamic potential
are given by the previous results (\ref{eff:bogo}) and
(\ref{regularized}) with this replacement.  

The momentum density, or equivalently, the mass current ${\bf g}$ of the
system is obtained in this approximation by differentiating the
effective potential with respect to $-{\bf u}$.  Show that 
\begin{equation} \label{bec:j}
{\bf g} = \bar{\rho}_{\rm s} {\bf v}_{\rm s} ,
\end{equation}
with $\bar{\rho}_{\rm s} = m \bar{n}$ the superfluid mass density.}
\vskip2pt \hrule  \vskip1pt \hrule \vskip4pt
\subsection{Impurities}
One of the ways to trigger a superconductor-insulator transition is to
change the amount of impurities.  This means that, {\it e.g.}, the
correlation length $\xi$ diverges as $|\hat{\alpha}^* -
\hat{\alpha}|^{-\nu}$ when the parameter $\hat{\alpha}$ characterizing
the impurities approaches the critical value $\hat{\alpha}^*$.

To account for impurities, we include a term \cite{Ma}
\begin{equation} \label{Dirt:dis}
{\cal L}_{\alpha} = \psi({\bf x}) \, |\phi(x)|^2
\end{equation} 
in the bosonic theory (\ref{bec:Lagr}), where $\psi({\bf x})$ is a
space-dependent random field with a Gaussian distribution
\begin{equation} \label{random} 
P[\psi] = \exp \left[-\frac{1}{\alpha} \int_{\bf x} \, \psi^2({\bf x})
\right],
\end{equation}
characterized by the impurity strength $\alpha$ ($\hat{\alpha}$ alluded
to above is a rescaled version of $\alpha$).  Notice that the random
field does not depend on time.  This is because it is introduced to
mimic impurities, which are randomly distributed in space, not in time.

We shall treat the impurities in the so-called quenched approximation
\cite{Ma}, where the average of an observable $O(\phi^*,\phi)$ is
obtained as follows
\begin{equation} 
\langle O(\phi^*,\phi) \rangle = \int_\psi P[\psi] \, \langle
O(\phi^*,\phi) \rangle_\psi,
\end{equation} 
with $\langle O(\phi^*,\phi) \rangle_\psi$ indicating the grand-canonical
average for a given impurity configuration.  That is to say, first the
ensemble average is taken for fixed $\psi$, and only after that the
averaging over the random field is carried out.

In terms of the shifted field (\ref{bec:newfields}), the random term
(\ref{Dirt:dis}) becomes
\begin{equation}  
{\cal L}_{\alpha} = \psi({\bf x}) (|\bar{\phi}_0|^2 + |\tilde{\phi}|^2 +
\bar{\phi}_0 \tilde{\phi}^* + \bar{\phi}_0^* \tilde{\phi}  ).
\end{equation} 
The first two terms lead to an irrelevant change in the chemical
potential, so that only the last two terms need to be considered, which
can be cast in the matrix form
\begin{equation}
{\cal L}_{\alpha} = \psi({\bf x}) \, \bar{\Phi}_0^\dagger \tilde{\Phi},
\;\;\;\;\;\;\;
\bar{\Phi}_0 = \left(\begin{array}{l} \bar{\phi}_0 \\ \bar{\phi}_0^*
\end{array} \right). 
\end{equation}
 
The integral over $\tilde{\Phi}$ is Gaussian in the Bogoliubov
approximation and therefore easily performed to yield an additional term
to the effective action
\begin{equation} 
S_{\alpha} = -\frac{1}{2} \int_{x,y} \psi({\bf x}) \bar{\Phi}_0^\dagger
\, G_0(x-y) \bar{\Phi}_0 \psi({\bf y}),
\end{equation} 
where the correlation function $G_0$ is the inverse of the matrix $M_0$
introduced in Eq.\ (\ref{bec:M}).  To proceed, we pass to the Fourier
representation:
\begin{eqnarray} 
G_0(x-y) &=& \int_k {\rm e}^{-i k \cdot (x-y)} \, G_0(k) \\ \psi({\bf
x}) &=& \int_{\bf k} {\rm e}^{i {\bf k} \cdot {\bf x}} \psi({\bf k}).
\end{eqnarray} 
The contribution to the effective action then appears in the form
\begin{equation} \label{S_d}
S_{\alpha} = -\frac{1}{2} \int_{\bf k} |\psi({\bf k})|^2
\bar{\Phi}_0^\dagger G(0,{\bf k}) \bar{\Phi}_0.
\end{equation} 
Since the random field is Gaussian distributed, the average over this
field representing quenched impurities yields:
\begin{equation} 
\langle |\psi({\bf k})|^2 \rangle = \tfrac{1}{2} \Omega \alpha,
\end{equation} 
with $\Omega$ the volume of the system.  The remaining integral over the
loop momentum in Eq.\ (\ref{S_d}) is readily carried out to yield in
arbitrary space dimensions the contribution to the Lagrangian
\begin{equation} \label{L_D}
\langle {\cal L}_\alpha \rangle = \frac{1}{2} \Gamma(1-d/2)
\left(\frac{m}{2 \pi} \right)^{d/2} |\bar{\phi}_0|^2 (6 \lambda_{\rm B}
|\bar{\phi}_0|^2 - \mu_{\rm B})^{d/2-1} \alpha. 
\end{equation} 
The divergence in the limit $d \to 2$ shows that also in the presence
of impurities, the two-dimensional case is special.  This
expression can be used to obtain the additional depletion due to
impurities.  To this end, we differentiate it with respect to the
chemical potential, giving \cite{GPS,pla}
\begin{equation} 
\bar{n}_\alpha = \frac{\partial \langle {\cal L}_\alpha
\rangle}{\partial \mu_{\rm B}} =
\frac{2^{d/2-5}\Gamma(2-d/2)}{\pi^{d/2}} m^{d/2} \lambda_{\rm B}^{d/2-2}
\bar{n}_0^{d/2-1} \alpha,
\end{equation}   
where we recall that $\bar{n}_0= |\bar{\phi}_0|^2$ denotes the
(constant) number density of particles residing in the condensate.
Because this contribution is positive, it amounts to an additional
depletion of the condensate.  The divergence in the limit $\lambda_{\rm
B} \rightarrow 0$ for $d <4$ signals the collapse of the system with
impurities when the interparticle repulsion is removed.

To determine the superfluid mass density $\bar{\rho}_{\rm s}$ in the
presence of impurities, we replace, as in the last exercise of Sec.\
\ref{sec:beyond}, $\mu_{\rm B}$ with $\mu_{\rm eff}$ defined in Eq.\
(\ref{bec:mureplacement}) and $i\partial_0$ with $i\partial_0 - ({\bf u}
- {\bf v}_{\rm s}) \cdot (-i \nabla)$ in the contribution (\ref{S_d}) to
the effective action.  Differentiating it with respect to the externally
imposed velocity, $-{\bf u}$, we find to linear order in the difference
${\bf u}- {\bf v}_{\rm s}$:
\begin{equation} 
{\bf g} = \bar{\rho}_{\rm s} {\bf v}_{\rm s} + \bar{\rho}_{\rm n} {\bf u},
\end{equation} 
with the superfluid and normal mass density \cite{pla}
\begin{equation} 
\bar{\rho}_{\rm s} = m\left(\bar{n} - \frac{4}{d} \bar{n}_\alpha
\right), \;\;\;\; \bar{\rho}_{\rm n} = \frac{4}{d} m \bar{n}_\alpha.
\end{equation}  
As expected, $\bar{\rho}_{\rm s} \neq m\bar{n}$ in the presence of
impurities.  Moreover, the normal mass density is a factor $4/d$ larger
than the mass density $m\bar{n}_\alpha$ knocked out of the condensate by
the impurities.  For $d=3$ this gives the factor $\tfrac{4}{3}$ first
found in Ref.\ \cite{HuMe}.  As argued there, this indicates that part
of the zero-momentum state belongs not to the condensate, but to the
normal fluid.  Being trapped by the impurities, the fraction $(4-d)/d
\times \bar{n}_\alpha$ of the zero-momentum state is localized.  

This is an important conclusion as it shows that the phenomenon of
localization can be accounted for in the Bogoliubov theory of
superfluidity by including a random field, without necessarily
destroying that state.
\section{Phase-only theory}
In this section we show that the Bogoliubov theory, which we obtained in
the composite boson limit after integrating out the fermionic degrees of
freedom from the theory of superconductivity, contains only one degree
of freedom, viz.\ the phase of the order parameter.  Physically, it
describes the Goldstone mode of the spontaneously broken global U(1)
symmetry.  In the context of superconductivity, this mode is called
Anderson-Bogoliubov mode.  The Bogoliubov theory may therefore, at least
in the superfluid state, be represented by a phase-only effective
theory.  We continue to account for the $1/r$ Coulomb potential in the
effective theory and give general scaling arguments for the physical
quantities represented by that theory.
\subsection{Derivation}
It was first shown by Beliaev \cite{Beliaev} that the gaplessness of the
single-particle spectrum first found by Bogoliubov at the classical
level persists at the one-loop order and later proven by Hugenholtz and
Pines \cite{HP} to hold to all orders in perturbation theory.  In fact,
as was proven by Gavoret and Nozi\`eres \cite{GN}, the Bogoliubov
spectrum is identical to that of the Goldstone mode accompanying the
spontaneous breakdown of the global U(1) symmetry, thus explaining its
gaplessness.

Also from the perspective of degrees of freedom, this conclusion makes
sense.  Although the normal phase is described by a complex
$\phi$-field, having two components, it contains only one degree of
freedom \cite{Leutwyler}.  This is because the energy $E({\bf k}) \sim
{\bf k}^2$ is always positive.  As a result, only positive energies
appear in the Fourier decomposition of the field, and one needs---as is
well known from standard quantum mechanics---a complex field to describe
a single spinless particle.  In the superfluid phase, on the other hand,
where $E^2({\bf k}) \sim {\bf k}^2$, the Fourier decomposition contains
positive as well as negative energies so that a single real field
suffices to describe this mode.  In other words, although the number of
fields is different, the number of degrees of freedom is the same in
both phases.  This implies that the superfluid state can be described by
a phase-only theory as it captures all the degrees of freedom, ignoring
vortices for the moment which are easily incorporated in the theory as
will be discussed in the next section.

To obtain the phase-only theory, we set, {\it cf.} Eq.\
(\ref{bec:newfields})
\begin{equation}
\phi(x) = {\rm e}^{i \varphi(x)} \, (\bar{\phi}_0 + \tilde{\phi}),
\end{equation}
with $\varphi(x)$ a background field representing the Goldstone mode
accompanying the spontaneous symmetry breakdown of the global U(1)
symmetry.  Inserting this in the Lagrangian (\ref{bec:Lagr}) and
expanding it, we obtain
\begin{equation} \label{quick}
{\cal L}^{(2)} = - {\cal V}_0 - |\bar{\phi}_0|^2 U - U (\bar{\phi}_0
\tilde{\phi}^* + \bar{\phi}_0^* \tilde{\phi} ) - \lambda_{\rm B}
|\bar{\phi}_0|^2 (\bar{\phi}_0 \tilde{\phi}^* + \bar{\phi}_0^*
\tilde{\phi} )^2,
\end{equation}
where the field $U(x)$ stands for the combination
\begin{equation}  \label{eff:U}
U(x) = \partial_0 \varphi(x) + \frac{1}{2m} [\nabla \varphi(x)]^2.
\end{equation}
In deriving Eq.\ (\ref{quick}), we used the mean-field equation
$\mu_{\rm B} = 2 \lambda_{\rm B} |\bar{\phi}_0|^2$.  We continue to
integrate out the tilde field (which is tantamount to substituting
its field equation back into the Lagrangian) to obtain the phase-only
theory
\begin{equation} \label{eff:quick}
{\cal L}_{\rm eff} = - \bar{n} U(x) + \frac{1}{4} U(x)
\frac{1}{\lambda_{\rm B}} U(x),
\end{equation}
where we ignored the irrelevant constant term ${\cal V}_0$ and
substituted $|\bar{\phi}_0|^2 (= \bar{n}_0) = \bar{n}$ to this order.
Using the mean-field equation again, we can write the coefficient of the
last term as:
\begin{equation} 
\frac{1}{4} \frac{1}{\lambda_{\rm B}} = \frac{1}{2} \frac{\bar{n}}{m c^2} =
\frac{1}{2} \bar{n}^2 \kappa, 
\end{equation} 
with $c$ the speed of sound introduced in Eq.\ (\ref{eff:micror}).
Standard thermodynamics relates $c$ to the compressibility $\kappa$ via
\begin{equation}
\kappa = \frac{1}{m \bar{n} c^2}.
\end{equation}
The phase-only theory (\ref{eff:quick}) can thus be cast in the
equivalent form
\begin{equation} \label{eff:Leff}
{\cal L}_{{\rm eff}} = -\bar{n}\left[\partial_{0}\varphi + \frac{1}{2m}(
{\bf \nabla} \varphi)^{2} \right] + \frac{1}{2} \bar{n}^2 \kappa
\left[\partial_{0}\varphi + \frac{1}{2m}( {\bf
\nabla}\varphi)^{2}\right]^{2},
\end{equation}
which turns out to be exact \cite{effbos}.  

The theory describes a sound wave, with the dimensionless phase field
$\varphi$ representing the Goldstone mode of the spontaneously broken
global U(1) symmetry.  It has the gapless spectrum  $E^2({\bf
k}) = c^2 {\bf k}^2$.  The effective theory gives, ignoring vortices for
the moment, a complete description of the superfluid at low energies and
small momenta.  When one goes to higher energies and momenta, additional
terms with higher-order derivatives should be included in the effective
theory, but it remains a phase-only theory.

\subsection{Coulomb potential}
It is straightforward to generalize the result (\ref{eff:quick}) to include
long-ranged interactions.  A case of particular interest to us is the
3-dimensional Coulomb potential
\begin{equation}
V({\bf x}) = \frac{q^2}{|{\bf x}|},
\end{equation}
whose Fourier transform in $d$ space dimensions reads
\begin{equation}
V({\bf k}) = 2^{d-1} \pi^{(d-1)/2} \Gamma[\tfrac{1}{2}(d-1)]
\frac{q^2}{|{\bf k}|^{d-1}}.
\end{equation}
Here, $q$ stands for the electric charge, which in the case of Cooper
pairs is twice the electron charge.  The simple contact interaction
$L_{\rm int} = - \lambda_{\rm B}\int_{\bf x} |\phi(x)|^4$ in Eq.\
(\ref{bec:Lagr}) is now replaced by
\begin{equation}
L_{\rm int} = - \frac{1}{2} \int_{{\bf x}, {\bf y}} |\phi(t,{\bf x})|^2
V({\bf x} - {\bf y}) |\phi(t,{\bf y})|^2.
\end{equation}
The rationale for using the 3-dimensional Coulomb potential, even
when considering charges confined to move in a lower dimensional space,
is that the electromagnetic interaction remains 3-dimensional.  The
effective theory then becomes after passing over to the Fourier
representation
\begin{equation}  \label{effCoul}
{\cal L}_{\rm eff} = - \bar{n} U(k)  + \frac{1}{2} U(k_0,{\bf k})
\frac{1}{V({\bf k})} U(k_0,-{\bf k}),
\end{equation}
and leads to the spectrum
\begin{equation}
E^2({\bf k}) = 2^{d-1} \pi^{(d-1)/2} \Gamma [\tfrac{1}{2}(d-1)]
\frac{\bar{n} q^2}{m} |{\bf k}|^{3-d}.
\end{equation}
For $d=3$, this yields the famous plasma mode, with an energy gap given
by the plasma frequency $\omega_{\rm p}^2 = 4 \pi \bar{n} q^2/m$.  For
$d=2$ on the other hand, the spectrum behaves as $E({\bf k}) \propto
\sqrt{|{\bf k}|}$, implying that the mode it describes is much harder
that the sound wave with the spectrum $E({\bf k}) \propto |{\bf k}|$
obtained for the system without the $1/r$ Coulomb interaction included.

To appreciate under which circumstances the Coulomb interaction becomes
important, we note that for electronic systems $1/|{\bf x}| \sim k_{\rm
F}$ for dimensional reasons and the fermion number density $\bar{n} \sim
k_{\rm F}^d$, where $k_{\rm F}$ is the Fermi momentum.  The ratio of the
Coulomb interaction energy to the Fermi energy $\epsilon_{\rm F} =
k_{\rm F}^2/2m$ is therefore proportional to $\bar{n}^{-1/d}$.  This
means that the lower the electron number density is, the more important
the Coulomb interaction becomes.
\subsection{Hyperscaling}
Let us consider the two terms in the effective theory (\ref{eff:Leff})
quadratic in the Goldstone field $\varphi$ and write them in the most
general form \cite{FF} 
\begin{equation} \label{general}
{\cal L}_{\rm eff}^{(2)} = - \frac{1}{2} \frac{\rho_{\rm s}}{m^2}
(\nabla \varphi)^2 + \frac{1}{2} \bar{n}^2 \kappa (\partial_0
\varphi)^2.
\end{equation}
The coefficient $\rho_{\rm s}$ is the superfluid mass density, which is,
as we saw in the previous section, a response function and in general
does not equal $m \bar{n}$.  The other coefficient,
\begin{equation}
\bar{n}^2 \kappa = \frac{\partial \bar{n}}{\partial \mu_{\rm B}} ,
\end{equation}
can be related to the (0,0)-component of the polarization tensor $\Pi_{0
0}$.  This can be understood by noting that an electromagnetic field is
included via the minimal substitution $\tilde{\partial}_\mu \to
\tilde{\partial}_\mu + q A_\mu$, with $A_\mu$ the electromagnetic vector
potential.  Since the polarization tensor (times $q^2$) is obtained by
differentiating the effective theory twice with respect to the vector
potential, we obtain
\begin{equation} \label{Pi} 
\lim_{|{\bf k}| \to 0} \Pi_{0 0} (0,{\bf k}) = \bar{n}^2 \kappa,
\end{equation} 
where, as is typical for response functions, the energy transfer is put
to zero before the momentum transfer ${\bf k}$ is.  Equation
(\ref{general}) leads to the general expression for the speed of sound
\begin{equation} \label{speed}
c^2 = \frac{\rho_{\rm s}}{m^2 \bar{n}^2 \kappa}.
\end{equation}

The singular behavior of the system close to the critical point is
encoded in the phase-only theory.  Simple dimensional analysis shows
that near the phase transition it scales as
\begin{equation}
{\cal L}_{\rm eff} \sim \xi^{-(d+z)},
\end{equation}
while 
\begin{equation} \label{scaling}
(\nabla \varphi)^2 \sim \xi^{-2}, \;\;\;\; (\partial_0 \varphi)^2 \sim
\xi_t^{-2} \sim \xi^{-2z},
\end{equation}
with $\xi_t$ the correlation time and $z$ the dynamic exponent.
Combining these hyperscaling arguments, and remembering that the mass
parameter is inessential with regards to the critical behavior, one
arrives \cite{FF} at the scaling laws for the two coefficients appearing
in the effective theory (\ref{general}):
\begin{equation} \label{hyperrho}
\rho_{\rm s} \sim \xi^{-(d+z-2)}, \;\;\;\; \kappa \sim \xi^{-(d-z)} .
\end{equation}
The first conclusion is consistent with the universal jump predicted by
Nelson and Kosterlitz \cite{NeKo} which corresponds to taking $z=0$ and
$d=2$. 

In the presence of impurities it is believed that the compressibility
stays finite at the critical point, implying $z=d$ \cite{FF}.  This
remarkably simple argument thus predicts an exact and nontrivial value
for the dynamic exponent.

Without impurities, the dynamic exponent is $z=2$ \cite{Uzunov}.  This
agrees with what one naively expects, given that in the nonrelativistic
theory (\ref{bec:Lagr}) we started with, one time derivative appears in
combination with two space derivatives, $i \partial_0 + \nabla^2/2m$.
This last argument should, however, be treated with care when applied to
the phase-only theory (\ref{eff:Leff}).  In that theory, the time and
space derivatives appear in a symmetrical form, yet $z$ is in general
not unity, as we just saw.  The difference is that in the effective
theory, the relative coefficient $c^2$ scales according to Eq.\
(\ref{speed}) with the scaling laws (\ref{hyperrho}) as
\begin{equation} 
c^2 \sim \xi^{2(1-z)},
\end{equation} 
while the relative coefficient $m$ in the microscopic theory does not
scale.  Incidentally, the (quantum) XY model has a dynamic exponent $z=1$,
so that $c$ in that case does not scale.

In experiments on charged systems, instead of the superfluid mass
density, usually the conductivity $\sigma$ is measured.  To see the
relation between the two, we introduce a vector potential in the effective
theory by replacing $\nabla \varphi$ with $\nabla \varphi - q {\bf A}$
in Eq.\ (\ref{general}), and allow the superfluid mass density to vary
in space and time.  The term in the action quadratic in ${\bf A}$ then
becomes after passing to the Fourier representation
\begin{equation}
S_\sigma = - \frac{1}{2} \frac{q^2}{m^2} \int_k {\bf A}(-k) \rho_{\rm s}
(k) {\bf A}(k).
\end{equation}
The electromagnetic current,
\begin{equation}
{\bf j}(k) = \frac{\delta S_\sigma}{\delta {\bf A}(-k)}
\end{equation}
obtained from this action can be written as
\begin{equation}
{\bf j}(k) = \sigma(k) {\bf E}(k),
\end{equation}
with the conductivity
\begin{equation} \label{conductivity}
\sigma(k) = i \frac{q^2}{m^2} \frac{\rho_{\rm s}(k)}{k_0}
\end{equation}
essentially given by the superfluid mass density.  So if we know the
scaling of the electric charge, we can determine the scaling of the
conductivity.

With the $1/r$ Coulomb potential included, the quadratic terms in the
effective theory (\ref{effCoul}) may, after passing to the Fourier
representation, be cast in the general form
\begin{equation} \label{L2}
{\cal L}_{\rm eff}^{(2)} = \frac{1}{2} \left(\frac{\rho_{\rm s}}{m^2}
{\bf k}^2 - \frac{|{\bf k}|^{d-1}}{q'^{\scriptstyle{2}}} k_0^2\right)
|\varphi(k)|^2,
\end{equation}
where $q'$ is the redefined charge parameter
\begin{equation}
q'^{\scriptstyle{2}} = 2^{d-1} \pi^{(d-1)/2}
\Gamma\left[\tfrac{1}{2}(d-1)\right] q^2.
\end{equation}
The charge is connected to the (0, 0)-component of the polarization
tensor via
\begin{equation}
q'^{\scriptstyle{2}} = \lim_{|{\bf k}| \rightarrow 0} \frac{|{\bf
k}|^{d-1}}{\Pi_{0 0} (0,{\bf k})} .
\end{equation}
A simple hyperscaling argument like the one given above for the case
without Coulomb interaction shows that near the transition, the
charge scales as \cite{FGG}
\begin{equation} \label{escaling}
q'^{\scriptstyle{2}} \sim \xi^{1-z},
\end{equation}
independent of the number of space dimensions $d$.  It then follows from
Eq.\ (\ref{conductivity}) that the conductivity scales as
\begin{equation} \label{sgeneral}
\sigma \sim \xi^{3 - (d+z)}.
\end{equation} 
\hrule  \vskip1pt \hrule \vskip4pt
{\footnotesize 
\noindent {\bf Exercise:} Give an alternative derivation of the result
(\ref{escaling}), using Eq.\ (\ref{Pi}).} 
\vskip2pt \hrule  \vskip1pt \hrule \vskip4pt

In the presence of random impurities, the charge is expected to be
finite at the transition, so that $z=1$ \cite{FGG}.  This is again an
exact result, which replaces the value $z=d$ of an impure system without
Coulomb interaction.  The prediction was first confirmed for impure
superconducting films \cite{HPsu1}, and has subsequently also been
observed in other 2-dimensional systems such 2-dimensional
Josephson-junction arrays \cite{Delft}, quantum Hall systems \cite{WET},
and 2-dimensional electron systems \cite{KSSMF}.  We will refer to a
quantum critical point with a $1/r$ Coulomb interaction as CQCP.  In the
vicinity of such a critical point, the conductivity scales as
\cite{Wegner}
\begin{equation} 
\sigma \sim \xi^{2 - d},
\end{equation} 
implying that in two space dimensions, the conductivity is a marginal
operator which remains finite at the CQCP.
\subsection{Scaling of magnetic vector potential}
\label{sec:magscale}
Let us finish this section by determining the scaling of the magnetic
vector potential.  We start with the observation that close to a CQCP,
the electric field $E$ scales as $E \sim \xi_t^{-1} \xi^{-1} \sim
\xi^{-(z+1)}$ (for a review, see Ref.\ \cite{SGCS}).
Thus conductivity measurements \cite{YaKa,KSSMF} close to a CQCP
collapse onto a single curve when plotted as function of the
dimensionless combination $|\delta|^{\nu (z+1)}/E$, where as before
$\delta=K-K_{\rm c})$ measures the distance from the critical
point $K_{\rm c}$, and $\nu$ is the correlation length exponent, $\xi
\sim |\delta|^{- \nu}$.  (For a field-controlled transition, $K$ stands
for the applied magnetic field, while for a density-controlled
transition it stands for the charge-carrier density.)  
The scaling of the electric field with the correlation length expresses
the more fundamental result that the anomalous scaling dimension $d_{\bf
A}$ of the magnetic vector potential ${\bf A}$ is unity, $d_{\bf A} =
1$.

Because the magnetic vector potential always appears in the
gauge-invariant combination $\nabla - q {\bf A}$, the anomalous scaling
dimension of the electric charge $q$ of the charge carriers times the
vector potential is unity too, $d_{q {\bf A}} = 1$.  Writing the
anomalous scaling dimension of the vector potential as a sum $d_{\bf A}
= d^0_{\bf A} + \tfrac{1}{2} \eta_{\bf A}$ of its canonical scaling
dimension $d^0_{\bf A} = \tfrac{1}{2} (d + z -2)$, obtained by simple
power counting, and (half) the critical exponent $\eta_{\bf A}$,
describing how the correlation function decays at the critical point, we
conclude that $d_q = d^0_q - \frac{1}{2}\eta_{\bf A}$.  Here, $d^0_q = 1
- d^0_{\bf A}$ stands for the canonical scaling dimension of the
electric charge.  Now, for a $1/r$ Coulomb potential, the charge scales
according to Eq.\ (\ref{escaling}) as $q^2 \sim \xi^{1-z}$ independent of
the number $d$ of space dimensions \cite{FGG}.  Combined with the
previous result, this fixes the value of the exponent $\eta_{\bf A}$ in
terms of the number of space dimensions and the dynamic exponent:
\begin{equation} \label{eta}
\eta_{\bf A}=5 -d - 2z.
\end{equation} 
Since in the presence of impurities, the electric charge is finite at a
CQCP, leading to $z=1$, it follows that $\eta_{\bf A}=1$ in two space
dimensions.

As we shall see in the next section, this exponent becomes important
when considering the interaction between vortices close to the CQCP.
\subsection{Experimental status}
For a critical discussion of the experimental status of the phase-only
theory, see Ref.\ \cite{LG}.  A more recent discussion can be found in
Ref.\ \cite{MG}.

According to the phase-only theory discussed here, no electronic
excitations exist in the critical region.  However, electron tunneling
measurements on superconducting films of varying thickness apparently
probed the energy gap of these excitations \cite{VDG}.  Moreover, the
gap was found to approach zero as the transition to the insulating state
is approached.  Similar experiments \cite{HCV} for the field-tuned
transition showed the presence of a large number of electronic
excitations near the Fermi energy, thus raising doubts about the
applicability of the phase-only theory.

Experimental support for the presence of electron pairs in the
insulating state comes from Hall effect studies on superconducting
films, which show two critical fields \cite{PHR}.  The lower critical
field is seen in the longitudinal resistance and is believed to mark the
superconductor-insulator transition.  The higher critical field is seen
in the transverse or Hall resistance and is believed to signal the
crossover from a bosonic to a fermionic insulator without pairing.  At
the higher critical field, the longitudinal resistance has its maximum.

The critical exponents determined in earlier experiments on the
superconductor-insulator transition \cite{HPsu1,YaKa} had the value $z =1$
for the dynamic exponent, in accord with the prediction in Ref.\
\cite{MPAFisher}, and $\nu =1.3$ for the correlation length exponent.
More recent studies \cite{MCMHG}, however, find agreement with these
results only for the transition tuned by changing the film
thickness.  For the field-tuned transition the value $z \nu = 0.7$
was found instead, which is about half the value one expects.  The cause
for this discrepancy is not clear.  It implies that, contrary to common
believe, the critical exponents depend on how the phase transition is
crossed, by tuning the field or the film thickness.

Clearly, more experimental and theoretical studies are required to fully
understand the superconductor-insulator transition, and to establish to
what extend the phase-only theory is applicable.

\section{Duality}
One of the most intriguing results found in experiments on quantum phase
transitions in superconducting films, as well as in 2-dimensional
Josephson-junction arrays \cite{Delft}, quantum Hall systems
\cite{IVQH}, and 2-dimensional electron systems \cite{KSSMF} is the
striking similarity in the current-voltage ($I$-$V$) characteristics on
both sides of the transition.  By interchanging the $I$ and $V$ axes in
one phase, an $I$-$V$ characteristic of that phase at a given value of
the applied magnetic field (in superconducting films, 2-dimensional
Josephson-junction arrays, and quantum Hall systems) or charge carrier
density (in 2-dimensional electron systems) can be mapped onto an
$I$-$V$ characteristic of the other phase at a different value of the
magnetic field or charge-carrier density.  This reflection symmetry
hints at a deep connection between the conduction mechanisms in the two
phases that can be understood by invoking a duality transformation
\cite{MPAFisher,WeZe}.  Whereas the conducting phase is most succinctly
described in terms of charge carriers of the system, the insulating
phase is best formulated in terms of vortices.  At zero temperature,
these topological defects should, just like the charge carriers, be
thought of as quantum point particles.  The duality transformation links
the two descriptions, which turn out to be very similar.
\subsection{Vortices}
\label{sec:Vortices}
Let us now include vortices in the phase-only theory.  This is achieved
by introducing the so-called plastic field $\varphi_\mu^{\rm P}$
via the minimal substitution $\tilde{\partial}_\mu \varphi \to
\tilde{\partial}_\mu \varphi + \varphi_\mu^{\rm P}$ \cite{GFCM}.
The plastic field is defined such that its curl gives a delta function
at the location of the vortices.  Specifically, in two space dimensions,
where vortices are point objects, located at the positions ${\bf
x}^\alpha$ say:
\begin{equation} \label{th2d}
\nabla \times \bbox{\varphi}^{\rm P} = -2 \pi \sum_\alpha \delta({\bf x}
- {\bf x}^\alpha),
\end{equation} 
while in three dimensions, where vortices are line objects, located along
the curves $C_\alpha$ say:
\begin{equation} 
\nabla \times \bbox{\varphi}^{\rm P} = -2 \pi \sum_\alpha
\int_{C_\alpha} \mbox{d} {\bf x}^\alpha \, \delta({\bf x} - {\bf
x}^\alpha).
\end{equation} 

Let us concentrate on static phenomena so that we can ignore the time
derivatives in the effective theory (\ref{general}).  When besides
vortices also the magnetic vector potential is included, the effective
theory becomes in three dimensions
\begin{equation} \label{starting}
{\cal L}_{\rm eff}^{(2)} = - \frac{1}{2} \frac{\rho_{\rm s}}{m^2}
(\nabla \varphi - \bbox{\varphi}^{\rm P} - q {\bf A})^2 - \frac{1}{2}
(\nabla \times {\bf A})^2,
\end{equation}
or after the canonical transformation $q {\bf A} \to q {\bf A} -
\bbox{\varphi}^{\rm P}$: 
\begin{equation} \label{dressed}
{\cal L}_{\rm eff}^{(2)} = - \frac{1}{2} \frac{\rho_{\rm s}}{m^2}
(\nabla \varphi - q {\bf A})^2 - \frac{1}{2} (\nabla \times {\bf A}-
{\bf B}^{\rm P})^2 ,
\end{equation}
where the plastic field ${\bf B}^{\rm P}$ stands for
\begin{equation} 
{\bf B}^{\rm P} = - \Phi_0 \sum_{\alpha}
\int_{C_\alpha} \mbox{d} {\bf x}^\alpha \, \delta({\bf x} - {\bf
x}^\alpha),
\end{equation} 
with $\Phi_0 = 2 \pi/q$ the magnetic flux quantum in units where the
speed of light and Planck's constant $\hbar$ are set to unity.  [In two
dimensions, this plastic field is a scalar and stands for
\begin{equation}    \label{BP}
B^{\rm P} = - \Phi_0 \sum_{\alpha} \delta({\bf x} - {\bf x}^\alpha),
\end{equation} 
as follows from Eq.\ (\ref{th2d}).]

After integrating out the phase field $\varphi$ in Eq.\ (\ref{dressed}),
we obtain the magnetic part of the effective action $S_{\rm mag}$.
Written as a functional integral over the magnetic vector potential, it
reads in the Coulomb gauge $\nabla \cdot {\bf A}=0$
\begin{equation} \label{mag}
{\rm e}^{i S_{\rm mag}} = \int_{\bf A} \, \exp \left\{i
\int_x \left[-\frac{1}{2}(\nabla \times {\bf A} - {\bf B}^{\rm
P})^2 - \frac{1}{2} \frac{1}{\lambda^2}{\bf A}^2 \right]\right\},
\end{equation} 
with $\lambda$ the magnetic penetration depth.  The mass term, with
$\lambda^{-2} = q^2\rho_{\rm s}/m^2$, is generated through the
Anderson-Higgs mechanism in the process of integrating out the phase
mode $\varphi$.  

With this construction, we can now calculate the interaction between two
vortices.  To facilitate the calculation in the case of a
superconducting film below, we linearize the first term in Eq.\
(\ref{mag}) by introducing an auxiliary field $\tilde{\bf h}$ via a
Hubbard-Stratonovich transformation:
\begin{equation} 
- \tfrac{1}{2} (\nabla \times {\bf A} - {\bf B}^{\rm P})^2 \to i (\nabla
\times {\bf A} - {\bf B}^{\rm P}) \cdot \tilde{\bf h} - \tfrac{1}{2}
\tilde{\bf h}^2.
\end{equation} 
The original form is regained after integrating out the auxiliary field
again.  After integrating out the magnetic vector potential, we arrive
at a form appropriate for a dual description in terms of magnetic
vortices rather than electric charges \cite{KKS}
\begin{equation} \label{3d}
{\rm e}^{i S_{\rm mag}} = \int_{\tilde{\bf h}} \, \exp\left\{i
\int_x \left[-\tfrac{1}{2} \lambda^2 (\nabla \times \tilde{\bf h})^2 -
\tfrac{1}{2}\tilde{\bf h}^2 - i \tilde{\bf h} \cdot {\bf B}^{\rm P}
\right] \right\}.
\end{equation} 
Physically, $\tilde{\bf h}$, which satisfies the condition $\nabla \cdot
\tilde{\bf h}=0$, represents ($i$ times) the fluctuating local
induction.  The vortices described by ${\bf B}^{\rm P}$ couple to
$\tilde{\bf h}$ with a coupling constant $g=\Phi_0/\lambda$ independent
of the electric charge.  Observe the close similarity between the
original (\ref{mag}) and the dual form (\ref{3d}).  This becomes even
more so when an external electric current ${\bf j}^{\rm P}$ is coupled
to the ${\bf A}$ field by including a term $- {\bf A} \cdot {\bf j}^{\rm
P}$ in Eq.\ (\ref{mag}), and ${\bf B}^{\rm P}$ describing the vortices
is set to zero there.

Finally, also integrating out the local induction, one obtains the
well-known Biot-Savart law for the interaction potential $S_{\rm mag} =
- \int_t V$ between two static vortices in a bulk superconductor
\cite{deGennes},
\begin{eqnarray} \label{V3d}
V(r) &=& \frac{1}{2\lambda^2} \int_{{\bf x},{\bf y}} B_i^{\rm P}({\bf
x}) G({\bf x} - {\bf y}) B_i^{\rm P}({\bf y}) \nonumber \\ &=&
\frac{g^2}{4 \pi} \int_{C_1} \int_{C_2} \mbox{d} {\bf l}^1 \cdot
\mbox{d} {\bf l}^2 \; \frac{{\rm e}^{-R/\lambda}}{R} \nonumber \\ &=& -
\frac{g^2}{2 \pi} L \left[\ln(r/2\lambda) + \gamma \right] + {\cal
O}(r/\lambda)^2 ,
\end{eqnarray} 
where we ignored the self-interaction.  In Eq.\ (\ref{V3d}), $G({\bf
x})$ is the vortex-vortex correlation function with Fourier transform
$G({\bf k})=1/({\bf k}^2 + \lambda^{-2} )$, $R$ denotes the distance
between the differential lengths $\mbox{d} {\bf l}^1$ and $\mbox{d} {\bf
l}^2$, $L$ is the length of each of the two vortices, and $\gamma$ is
Euler's constant.  For distances smaller than the magnetic penetration
depth, which is the length scale for variations in the electric current
and the magnetic field, the interaction is logarithmic as in a
superfluid.  If the system size is smaller than $\lambda$, it will
replace the penetration depth as infra-red cutoff in the logarithm, and
there will be no reference to the electric charge anymore.

To describe magnetic vortices in a film of thickness $w$ \cite{Pearl},
the bulk result (\ref{3d}) has to be adjusted in two ways to account for
the fact that both the vortices and the screening currents, which
produce the second term in (\ref{3d}), are confined to the plane.  This
is achieved by including a Dirac delta function $w \delta(x_3)$ in the
second and third term.  Instead of Eq.\ (\ref{3d}), we then arrive at
the interaction potential \cite{Pearl,deGennes}
\begin{eqnarray} \label{V2d}
V_\perp(r) &=& \frac{1}{2 \lambda_\perp} \int_{{\bf x}_\perp, {\bf y}_\perp}
B_\perp^{\rm P}({\bf x}_\perp) G_\perp({\bf x}_\perp
- {\bf y}_\perp) B_\perp^{\rm P}({\bf y}_\perp) \nonumber \\ &=&
-\frac{g_\perp^2}{2 \pi} \left[ \ln(r/4 \lambda_\perp) + \gamma\right] +
{\cal O}(r/\lambda_\perp)^2,
\end{eqnarray} 
where $B_\perp^{\rm P} = - \Phi_0 \sum_{\alpha} \delta({\bf x}_\perp -
{\bf x}_\perp^\alpha)$ describes the vortices in the film with
coordinates ${\bf x}_\perp$, $\lambda_\perp = \lambda^2/w$ is the
transverse magnetic penetration depth, $g_\perp^2 =
\Phi_0^2/\lambda_\perp$ the coupling constant squared, and
\begin{eqnarray} 
G_\perp({\bf x}_\perp) &=& \int_{x_3} \, G_\perp({\bf x}_\perp, x_3)
\nonumber \\ &=& \int_{{\bf k}_\perp} \, {\rm e}^{- i {\bf k}_\perp
\cdot {\bf x}_\perp} G_\perp({\bf k}_\perp,0),
\end{eqnarray} 
with $G_\perp({\bf k}_\perp,0) = 2/ k_\perp (2 k_\perp
+\lambda_\perp^{-1})$.  For small distances, the interaction is seen to
be identical to that in a bulk superconductor \cite{Pearl}, and also to
that in a superfluid film.  As in the bulk, the vortex coupling constant
$g_\perp$ in the film is independent of the electric charge.

The logarithmic interaction between vortices we found in Eq.\
(\ref{V2d}) appears to pose a severe problem to the duality picture we
alluded to in the introduction of this section as the charges interact
via a $1/r$ Coulomb potential.  The difference should spoil the
experimentally observed reflection symmetry in the $I$-$V$
characteristics.  However, it should be realized that the results
derived in this subsection are valid only in the mean-field region,
where $\eta_{\bf A}=0$.  In the critical region governed by a CQCP, the
value of this exponent was found in Sec.\ \ref{sec:magscale} to be
unity.  As we will now demonstrate, this leads to a qualitative change
in the interaction potential between two vortices from logarithmic in
the mean-field region to $1/r$ in the vicinity of the CQCP.

\subsection{Changing vortex interaction}
Close  to a  CQCP we  have to  include the  field  renormalization factor
$Z_{\bf  A}$   in  the  vortex-vortex   correlation  function  $G_\perp$
appearing in the expression (\ref{V2d}) for the vortex interaction.  It
then becomes
\begin{equation} \label{Gr} 
G_\perp({\bf k}_\perp,0) = \frac{2}{k_\perp } \frac{Z_{\bf
A}}{ 2 k_\perp +\lambda_\perp^{-1}},
\end{equation} 
with $Z_{\bf A} \sim k_\perp^{\eta_{\bf A}}$.  Because the magnetic
vector potential and the local induction renormalize in the same way,
their renormalization factor is identical.  Due to this extra factor,
the interaction between two vortices in the film takes the form of a
$1/r$ Coulomb potential \cite{PRL2k}
\begin{equation}  \label{renV}
V_\perp(r) = \frac{g_\perp^2}{2 \pi} \frac{a}{r},
\end{equation} 
where $a$ is some microscopic length scale which accompanies the
renormalization factor $Z_{\bf A}$ for dimensional reasons
\cite{Goldenfeld}.  

The absence of any reference to the electric charge in the renormalized
and bare interaction (at least for small enough systems) implies that
the same results should be derivable from our starting theory
(\ref{starting}) with $q$ set to zero.  By directly integrating out the
Anderson-Bogoliubov mode, and ignoring the momentum dependence of
$\rho_{\rm s}$, which is valid outside the critical region, one easily
reproduces the bare interaction potential (\ref{V2d}).  The renormalized
interaction (\ref{renV}) is obtained by realizing that according to Eq.\
(\ref{hyperrho}), $\rho_{\rm s} \sim k_\perp$ for $d=2$ and $z=1$.  In
other words, the extra factor of $k_\perp$ that came in via the
renormalization factor $Z_{\bf A}$ in our first calculation to produce
the $1/r$ potential, comes in via $\rho_{\rm s}$ here.

One might wonder if perhaps also the Coulomb interaction between
electric charges changes in the vicinity of a CQCP.  We do not expect
this to happen.  Since the $1/r$ Coulomb interaction is genuine
3-dimensional, this interaction cannot be affected too much by what
happens in the film, which constitutes a mere slice of 3-dimensional
space.  The reason that the interaction between vortices is susceptible
to the presence of a CQCP, is that this interaction is a result of
currents around the vortex cores which are confined to the plane.

A similar change in the $r$-dependence of the interaction between two
vortices upon entering a critical region has been observed numerically
in the 3-dimensional Ginzburg-Landau model \cite{OlTe}.  Near the
charged fixed point of that theory, $\eta_{\bf a}=1$ \cite{HeTe}, as in
our case.  

This is a very pleasing coincidence as the (2+1)-dimensional
Ginzburg-Landau model constitutes the dual formulation of the system.
\subsection{Dual Theory}
The {\it dynamics} of the charged degrees of freedom is described by the
effective Lagrangian (\ref{general}) with the speed of sound given by
Eq.\ (\ref{speed}).  In accord with the above findings, we have ignored
the coupling to the magnetic vector potential, so that the effective
theory essentially describes a superfluid.

In the dual formulation, the roles of charges and vortices are
interchanged.  And the Anderson-Bogoliubov mode mediating the
interaction between two vortices is represented as a photon associated
with a fictitious gauge field $a_\mu$, i.e., $\tilde{\partial}_\mu
\varphi \sim \epsilon_{\mu \nu \lambda} \tilde{\partial}_\nu a_\lambda$.
In 2+1 dimensions, this identification makes sense as a photon has only
one transverse direction and thus only one degree of freedom.  It
therefore represents the same number of degrees of freedom as does the
Anderson-Bogoliubov mode.  

The elementary excitations of the dual theory are the vortices,
described by a complex scalar field $\psi$.  Specifically, the dual
theory of Eq.\ (\ref{general}) turns out to be the Ginzburg-Landau model
\cite{dualGL,GFCM,WeZe,KKS}
\begin{equation} 
{\cal L}_{\rm dual} = -\tfrac{1}{4} f_{\mu \nu}^2 + |(\partial_\mu -i g
a_\mu) \psi|^2 - m_\psi^2 |\psi|^2 - u|\psi|^4,
\end{equation} 
with $f_{\mu \nu} = \tilde{\partial}_\mu a_\nu - \tilde{\partial}_\nu
a_\mu$, $m_\psi$ a mass parameter, and $u$ the strength of the
self-coupling.  Both the gauge part as well as the matter part of the
dual theory are of a relativistic form.  The gauge part is because the
effective theory (\ref{general}), obtained after ignoring nonlinear
terms, is Lorentz invariant.  The matter part is because vortices of
positive and negative circulation can annihilate, and can also be
created.  In this sense they behave as relativistic particles.  As was
pointed out in Ref.\ \cite{WeZe}, the speed of ``light'' in the gauge
and matter part need not to be identical and will in general differ.

The interaction potential (\ref{V2d}) between two external vortices is
now being interpreted as the 2-dimensional Coulomb potential between
charges.  The observation concerning the critical behavior of the
Ginzburg-Landau model implies that the qualitative change in $V(r)$
upon entering the critical region is properly represented in the dual
formulation.

Whereas in the conducting phase, the charges are condensed, in the
insulating phase, the vortices are condensed \cite{MPAFisher}.  In the
dual theory, the vortex condensate is represented by a nonzero
expectation value of the $\psi$ field, which in turn leads via the
Anderson-Higgs mechanism to a mass term for the gauge field $a_\mu$.
Because $(\epsilon_{\mu \nu \lambda} \tilde{\partial}_\nu a_\lambda)^2
\sim (\tilde{\partial}_\mu \varphi)^2$, the mass term $a_\mu^2$ with two
derivatives less implies that the Anderson-Bogoliubov mode has acquired
an energy gap.  That is to say, the phase where the vortices are
condensed is incompressible and indeed an insulator.  Since electric
charges are seen by the dual theory as flux quanta, they are expelled
from the system as long as the dual theory is in the Meissner state.
Above the critical field $h = \nabla_\perp \times {\bf a} = h_{c_1}$
they start penetrating the system and form an Abrikosov lattice.  In the
original formulation, this corresponds to a Wigner crystal of the
charges.  Finally, when more charges are added and the dual field
reaches the critical value $h_{c_2}$, the lattice melts and the charges
condense in the superfluid phase described by the effective theory
(\ref{general}).
\section*{Acknowledgments}
It is a pleasure to thank J. Spa{\l}ek and the other organizers of the
XL Cracow School of Theoretical Physics for inviting me to lecture at
the renowned summer school at Zakopane, Poland.  I am indebted to
M. Krusius for the hospitality at the Low Temperature Laboratory,
Helsinki University of Technology, Finland where the lectures were
prepared, and to B. Rosenstein for the hospitality at the Department of
Electrophysics, National Chiao Tung University, Hsinchu, Taiwan where
these notes were written.  I would also like to acknowledge helpful
discussions with G. Honig and M. Paalanen.

This work was funded in part by the EU sponsored programme Transfer and
Mobility of Researchers under contract No.\ ERBFMGECT980122, and by the
National Science Council (NCS) of Taiwan.

\end{document}